\documentclass{article}

  \usepackage[final]{tackling_climate_workshop_style}

\usepackage[utf8]{inputenc} 
\usepackage[T1]{fontenc}    
\usepackage{hyperref}       
\usepackage{url}            
\usepackage{booktabs}       
\usepackage{amsfonts}       
\usepackage{nicefrac}       
\usepackage{microtype}      
\usepackage{colortbl}
\usepackage{graphicx}
\usepackage{subcaption}
\usepackage{adjustbox} 
\usepackage{graphicx}
\graphicspath{{./}{figures/}} 

\newcommand{\bee}{\raisebox{-0.5ex}{\includegraphics[height=1.5em]{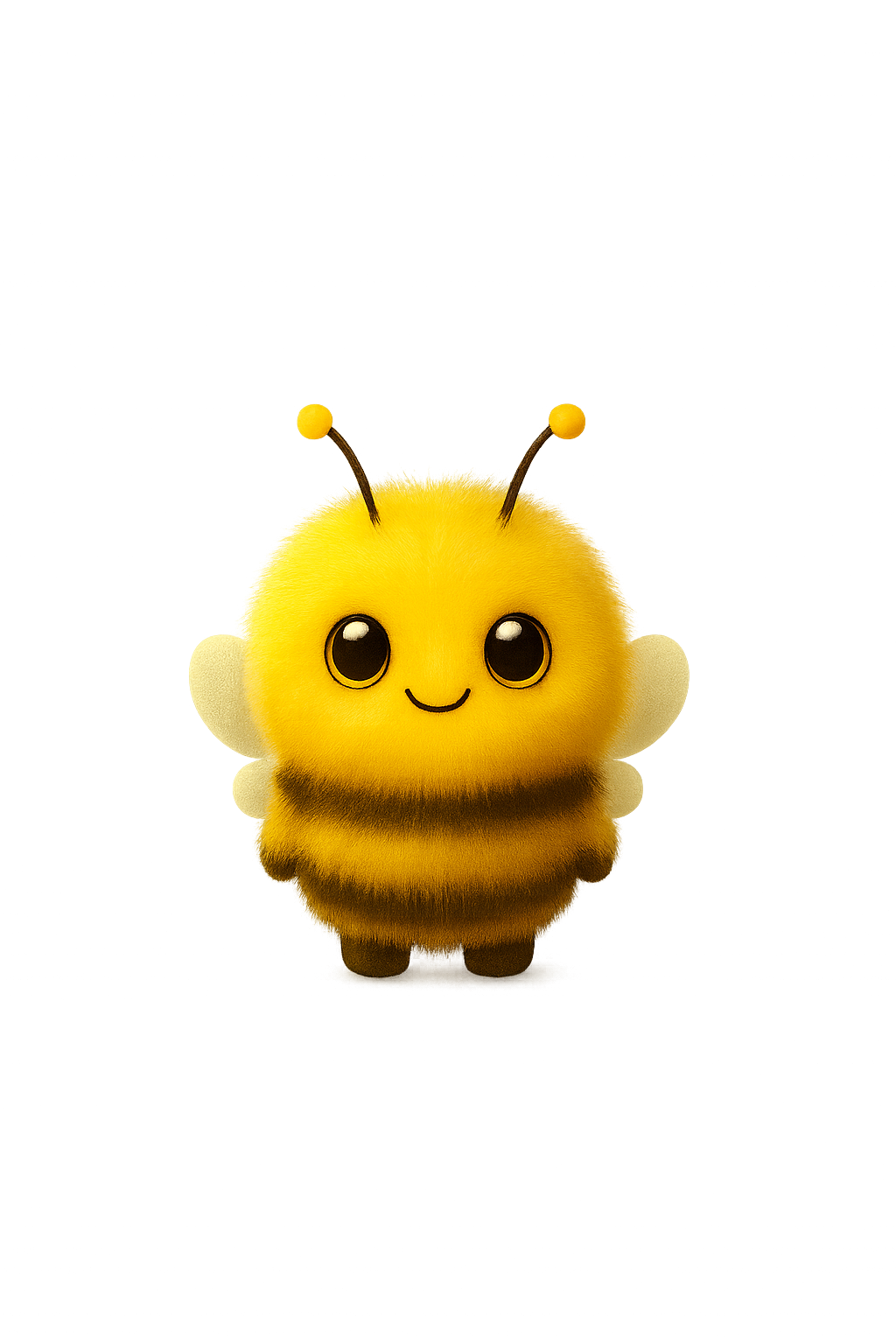}}}

\title{

 \bee Eco-Bee: A Personalised Multi-Modal Agent for Advancing Student Climate Awareness and Sustainable Behaviour in Campus Ecosystems
}

\author{%
  Caleb Adu\footnotemark[1] \\
  University of Hull \\
  Hull, United Kingdom \\
  The Spaceship Academy \\
  London, United Kingdom \\
  \And
  Neil Kapadia\footnotemark[1] \\
  City St George's, University of London \\
  London, United Kingdom \\
  The Spaceship Academy \\
  London, United Kingdom \\
  \And
  Binhe Liu\footnotemark[1] \\
  King's College London \\
  London, United Kingdom \\
  The Spaceship Academy \\
  London, United Kingdom \\
  \AND
  Jonathan Randall\footnotemark[1] \\
  Lancaster University \\
  Lancaster, United Kingdom \\
  The Spaceship Academy \\
  London, United Kingdom \\
  \And
  Sruthi Viswanathan \\
  University of Oxford \\
  Oxford, United Kingdom \\
  The Spaceship Academy \\
  London, United Kingdom \\
}

\begin{document}

\maketitle

\newcommand{\sruthi}[1]{{\color{magenta} {#1}}}

\footnotetext[1]{Equal contribution.}
\begin{abstract}
Universities are microcosms of urban ecosystems, with concentrated consumption patterns in food, transport, energy, and product usage. These environments not only contribute substantially to sustainability pressures but also provide a unique opportunity to advance sustainability education and behavioural change at scale. As in most sectors, digital sustainability initiatives within universities remain narrowly focused on carbon calculations, typically providing static feedback that limits opportunities for sustained behavioural change. To address this gap, we propose Eco-Bee, integrating large language models, a translation of the Planetary Boundaries framework (as Eco-Score), and a conversational agent that connects individual choices to environmental limits. Tailored for students at the cusp of lifelong habits, Eco-Bee delivers actionable insights, peer benchmarking, and gamified challenges to sustain engagement and drive measurable progress toward boundary-aligned living. In a pilot tested across multiple campus networks (n=52), 96\% of the student participants supported a campus-wide rollout and reported a clearer understanding of how daily behaviours collectively impact the planet’s limits. By embedding planetary science, behavioural reinforcement, and AI-driven personalisation into a single platform, Eco-Bee establishes a scalable foundation for climate-conscious universities and future AI-mediated sustainability infrastructures.

\end{abstract}

\section{Introduction And Related Work}
Individual actions alone cannot solve the climate crisis, but they are a critical starting point~\cite{stern2000role}. Embedding sustainable habits in daily life builds awareness that often extends into professional and societal choices~\cite{otto2020social}. Students, at the cusp of their careers, are uniquely positioned to internalise planetary-boundaries alignment ~\cite{rockstrom2009planetary} and extend it into future impact sectors. Cultivating awareness early among students can catalyse both personal and systemic change \cite{su17146324}.

Students today are increasingly aware of the climate crisis—and often experience climate anxiety as a result~\cite{hickman2021climate}. Tools such as UC Berkeley’s CoolClimate calculator estimate greenhouse gas (GHG) emissions across transport, energy, food, goods, and services and offer personalised footprints with peer benchmarking~\cite{coolclimate2025}. Behavioural science shows that social comparison feedback can reduce energy use by up to 7\%~\cite{ayres2013peer}, while gamified features like badges and leaderboards enhance engagement and autonomy in sustainability apps~\cite{boncu2022gameful}. Yet, many existing tools remain carbon-centric and static, lacking integration with multi-dimensional environmental metrics and student-contextualised behaviour support. This narrow focus, often termed “carbon tunnel vision,” risks obscuring broader environmental dynamics essential to sustainable decision making~\cite{deivanayagam2023breaking}.

To address this gap, we introduce \textbf{Eco-Bee}—a multi-modal AI agent uniting vision–language models, planetary-boundary-aligned impact metrics, and graph-based recommendations into a unified feedback loop. Eco-Bee enquires about student behaviours on campus, quantifies their impact across the nine planetary boundaries (Climate Change, Biosphere Integrity, Biogeochemical Flows, Land-System Change, Freshwater Use, Ocean Acidification, Atmospheric Aerosol Loading, Stratospheric Ozone Depletion, and Novel Entities), and delivers personalised, gamified recommendations to nurture actionable change. This novel fusion of behavioural science, AI, and boundary-aware feedback is designed to transform student climate concern into measurable, socially reinforced action. In the remainder of this proposal, we present our pilot implementation and testing of Eco-Bee, outline our roadmap, and discuss its potential impact in driving systemic, AI-mediated behavioural change for sustainable climate futures across universities and beyond.

\section{Project Eco-Bee}

\subsection{Planning and Co-Design}
The design of Eco-Bee (see Figure 1) was informed by a co-design process conducted in consultation with two experts, a climate innovation specialist and a university programmes director. This process ensured that the system aligned with both the scientific rigour of the planetary boundaries framework, climate innovation strategies, and the practical needs of students in campus ecosystems. Key design principles identified:
\begin{itemize}
    \item Provide \textit{personalised, multi-boundary feedback} rather than generic carbon-only metrics.
    \item Use \textit{gamified, socially reinforced engagement} to foster motivation while maintaining autonomy.
    \item Incorporate \textit{privacy by design}, with minimal and aggregated data collection.
    \item Build a \textit{scalable, modular architecture} to support rapid iteration and future integrations.
\end{itemize}


\subsection{System Architecture and Implementation}

We deployed a pilot version of Eco-Bee as a responsive, web-based platform deployed on Vercel, with a Next.js frontend and a Python \textit{FastAPI} service for scoring, recommendations, and leaderboard aggregation. Data are persisted in Supabase (managed Postgres) with Row Level Security (RLS) policies to protect user records and scores. The vision pipeline uses \textit{Mistral’s Pixtral-12B} multimodal large language model for image understanding (meals, clothing, product barcodes), with images processed in-session and only derived categories stored. Features of Eco-Bee include:
\begin{itemize}
  \item \textbf{Multi-modal intake service}: captures quiz responses, optional images, and barcodes; normalises inputs for downstream scoring.
  \item \textbf{Vision AI service}: calls Pixtral-12B for image classification and captions, returning sustainability-relevant labels.
\item \textbf{Scoring engine:} translates user-reported behaviours into planetary-boundary factors using a curated set of domain–boundary mappings (e.g., diet, mobility, fashion, and housing behaviours mapped to climate, biosphere, and freshwater pressures). It returns both boundary-level scores and a composite Eco-Score, normalised to a 0–100 scale (see Appendix~\ref{appendix:pseudocode} for detailed pseudocode).

\item \textbf{Recommendation engine:} a node2vec-based graph ranks feasible against each user’s lowest boundary scores to surface the most contextually relevant suggestions.

\item \textbf{Social and leaderboard service:} Aggregates anonymised Eco-Scores and boundary-level data to enable pesudonymised peer benchmarking and lightweight gamification, supporting social reinforcement.

  \item \textbf{Chatbot service}: open-ended conversation with Pixtral 12B, serves opportunities curated for the university context to make recommendations actionable.
  \item \textbf{Privacy layer}: pseudonymous IDs, RLS-guarded tables, transient image handling, and aggregate-only displays.
\end{itemize}


\begin{figure*}[t]
  \centering
  \begin{adjustbox}{max width=\textwidth,keepaspectratio}
    \begin{minipage}[c][0.40\textwidth][c]{\textwidth}
      \centering
      \begin{subfigure}[t]{0.16\textwidth} 
        \centering
        \includegraphics[width=\linewidth]{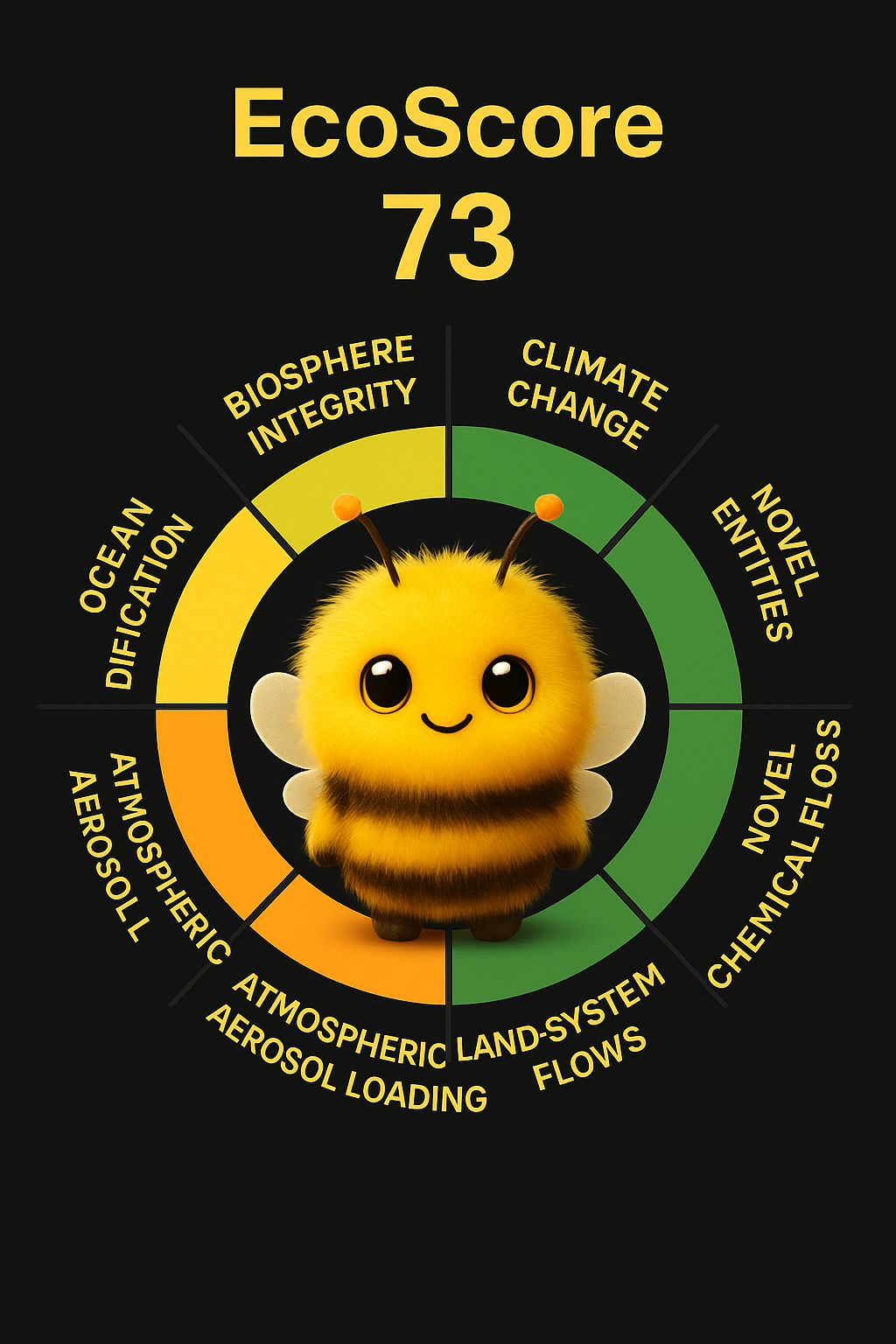}
      \end{subfigure}\hfill
      \begin{subfigure}[t]{0.32\textwidth}
        \centering
        \includegraphics[width=\linewidth]{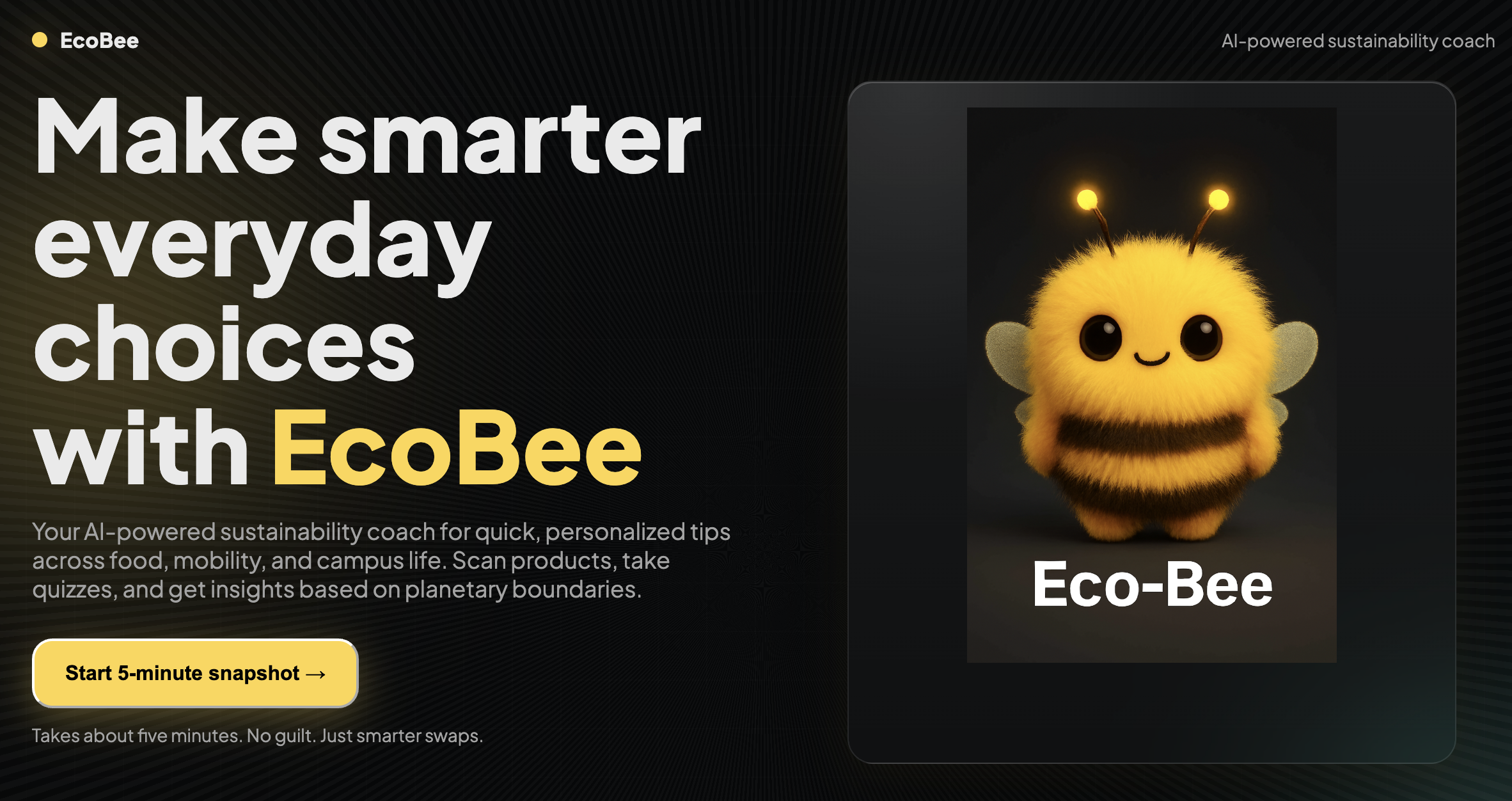}
      \end{subfigure}\hfill
      \begin{subfigure}[t]{0.46\textwidth}
        \centering
        \includegraphics[width=\linewidth]{ebsysarch.png}
      \end{subfigure}
    \end{minipage}
  \end{adjustbox}
  \caption{Overview of Eco-Bee: (Left) initial idea used for co-design interactions; (Centre) screenshot from the pilot's deployment on Vercel; (Right) backend system architecture.}
  \label{fig:ecobee-16x9}
  \vspace{-7pt}
\end{figure*}

\subsection{Pilot User Testing and Results}
\label{sec:analysis_results}
The initial pilot study was launched through targeted outreach to student networks across multiple channels, including university mailing lists, and groups on social media.  A total of \textbf{52 students} participated in the pilot, engaging with Eco-Bee to calculate their Eco-Scores and receive personalised recommendations. Among these, \textbf{26 students} completed a follow-up questionnaire providing structured feedback on usability, clarity, motivational effect, and perceived future use of Eco-Bee.

Analysis of the 26 survey responses indicated strong engagement and a positive reception. Students rated \textit{clarity of interaction} at 4.0/5, \textit{ease of navigation} at 4.4/5, and \textit{usefulness of recommendations} at 4.1/5. \textbf{96\%} expressed interest in a campus-wide rollout of Eco-Bee, and \textbf{80\%} showed willingness to engage with peer benchmarking features.

Qualitative feedback highlighted Eco-Bee’s ability to make the link between daily actions and planetary impacts more tangible. The Eco-Score breakdown helped students better understand sustainability trade-offs and motivated many to adopt lower-impact alternatives. Suggestions for refinement centred on improving the mobile interface, enhancing the gamification layer, and tailoring recommendations to local campus resources.

Composite Eco-Scores from the quiz (\(n=52\)) followed a unimodal distribution centred in the low-50s (mean = 50.9, SD = 7.7, range = 36--62). Per-boundary averages from the leaderboard (Top score = 62/100) were broadly consistent across domains, with Climate \(\approx 48.6\), Biosphere \(\approx 53.6\), and Biogeochemical, Freshwater, and Aerosols \(\approx 50\) (0--100 scale).

\section{Pathway to Climate Impact and Conclusion}
Eco-Bee charts a scalable pathway for AI-mediated societal climate action. Its planetary-boundaries-based Eco-Score and behavioural recommender can extend beyond campuses to municipal systems, corporate sustainability programmes, and a global student sustainability network. We plan to evolve Eco-Bee into a scalable platform that integrates adaptive recommendations, richer campus-specific datasets, and longitudinal analytics to drive sustained behaviour change and inform evidence-based sustainability policies. By making climate change actionable with AI, Eco-Bee has the potential to help keep communities and humanity within a safe operating space for the planet’s systems.

\begin{ack}

\end{ack}





\bibliographystyle{plain}
\bibliography{eco_bee_references}

\clearpage
\appendix
\section{Pseudocode for Quiz Flow and EcoScore Computation}
\label{appendix:pseudocode}

\subsection{Quiz Flow (Frontend and API)}
\begin{verbatim}
state Profile = {
  cohort, campus, faculty, pseudonym,
  quiz: { diet, mobility, fashion, housing, career_interest, ... },
  images?: [dataURL], barcodes?: [string]
}

onStartQuiz():
  show consent + PB explainer
  init empty Profile; start Stepper
...
POST /api/submit-score { pseudonym, campus, boundaries, composite, feedback }
  SUPABASE.insert("leaderboard", {pseudonym, campus, composite, boundary_scores: boundaries})
  if feedback: SUPABASE.insert("feedback", feedback)
  return {ok: true}
\end{verbatim}

\subsection{EcoScore Computation (Backend FastAPI)}
\begin{verbatim}
// --- factor tables (loaded at startup) ---
FACTORS = {
  food: CSV("factors_food.csv"),
  mobility: CSV("factors_mobility.csv"),
  ...
}

// --- main scoring ---
POST /api/score(payload):
  items = canonicalise_quiz(payload.quiz)
  if payload.labelled_items: items += payload.labelled_items
  ...
  return {boundaries: boundary_scores, composite, explanations}
\end{verbatim}

\end{document}